\documentclass[aps,prl,preprint,showpacs]{revtex4}
\usepackage{epsfig}
\usepackage{amssymb}

\begin{document}

\title{Motion of  a sphere in an oscillatory boundary layer: an optical tweezer based study }
\author{Prerna Sharma \footnote{email : prerna@tifr.res.in}, Shankar Ghosh and S. Bhattacharya}
\date{\today }

\begin{abstract}
The drag forces acting on a single polystyrene sphere in the
vicinity of an oscillating glass plate have been measured using an
optical tweezer. The phase of the sphere is found to be a sensitive
probe of the dynamics of the sphere. The evolution of the phase from
an inertially-coupled regime to a purely velocity-coupled regime is
explored. Moreover, the frequency dependent response is found to be
characteristic of a damped oscillator with an effective inertia
which is several orders of magnitude greater than that of the
particle.
\end{abstract}

\affiliation{Department of Condensed Matter Physics and Materials Science\\
Tata Institute of Fundamental Research\\
Homi Bhabha Road, Mumbai 400-005, India\\
}
\maketitle

\section{Introduction}

Study of the forces between two sliding bodies in the presence of an
intervening fluid is an important problem in fluid mechanics
\cite{landau}. In recent years, a variety of experimental and
theoretical/ computational tools have been brought to bear of this
central issue. Experimental studies of the ``simple'' case of
hydrodynamic drag forces acting on a single particle suspended in a
liquid near a plate executing sinusoidal oscillations are rare,
although there have been measurements on the lift forces acting on
spheres in oscillatory flows \cite{sleath}. Such measurements have
shown a highly non-monochromatic response of the lift force as a
function of the oscillating frequency. This problem is important in
understanding the role played by colloidal particles in a
lubricating liquid as well as the nature of interactions between the
colloidal particles and the solid interface formed say by the walls
of narrow channels, e.g., blood vessels.

In this paper we present a new technique based on an optical tweezer
that addresses an important aspect of this problem: how does one
probe the dynamical response of a solid particle in ``contact'' with
a substrate in the presence of a lubricating liquid, which requires
an elucidation of the notion of the ``contact'' itself. The study
reveals the importance of the phase response of the particle which
provides insight into the nature of the dynamics.

A schematic of the experimental setup is shown in figure 1 . A very
dilute monodisperse colloidal solution of polystyrene spheres of
radius(a) of 1$\mu $m (Alfa Aesar) was placed between two parallel
glass plates separated by a distance of 2mm. The top glass plate was
kept stationary. The bottom glass plate was attached to a $xyz$
piezo stage subjected to an oscillatory motion in the $y$ direction
with an amplitude $y_{po}$ and frequency $\omega $, i.e.,
$y_{p}=y_{po}\sin (\omega t)$, where $y_{p}$ is the instantaneous
displacement of the glass plate . A colloid particle was trapped
near the glass plate using an IR laser of wavelength 1064 nm focused
using a 100X oil immersion objective lens. A He-Ne laser of
wavelength 632.8 nm was used to track the position of the sphere.
The position of the sphere was detected using a quadrant photo diode
(UDT instruments). The positional response of the sphere was then
locked on to the drive signal of the piezo plate using a SRS 830
lock-in amplifier and both phase and amplitude of the locked in
signal were analyzed. The separation between the particle and the
oscillating glass plate was varied by moving the piezo stage in the
$z$ direction. The trap stiffness and the corner frequency of the
trap were analyzed using the power spectrum of the equilibrium
positional fluctuations of the trapped particle.

\section{Forces acting on the sphere due to oscillatory flow}

The two extreme scenarios possible in the problem are (i) when the
sphere is in physical contact with the bottom plate and (ii) when
the sphere is far away from the glass plate. In the first scenario
the coupling between the sphere and the bottom plate is mainly
inertial and this is reflected in an ``in-phase'' response of the
sphere with respect to the motion of the bottom plate.  When the
sphere is far away from the plate then the forces acting on the
sphere will be dominantly due to viscosity mediated shear stresses
and drag forces arising from the motion of the liquid. Therefore, in
this experiment we have two control parameters, i.e., the height of
the sphere from the plate and the velocity of the plate.

The motion of the liquid in presence of oscillatory shear is
governed by the Naiver Stokes equation $\rho\left( \frac{\partial
v}{\partial t}+(v.\nabla )v\right) =-\nabla p+\eta \nabla ^{2}v$,
where $v$ is the velocity, $\eta $ is viscosity ,$\nabla p$ is the
pressure gradient and $\rho $ is the density of the liquid. For the
present problem there is no pressure gradient, i.e., $\nabla p=0$,
and equation of continuity demands $\mathbf{\nabla \cdot \ v}=0$,
thus $(v.\nabla )v=0$. With the above assumptions the Naiver Stokes
equation can be written as $ \rho \frac{{\partial v_y }}{{\partial
t}} = \eta \frac{{\partial ^2 v_y }}{{\partial z^2 }} $. This has
traveling wave solutions of the form, $ v_y (z,t) = v_0 e^{i(kz -
\omega t)} $ with Stokes boundary layer, $\delta =\sqrt{\frac{2\eta
}{\omega\rho }}$ and $k=\pm \frac{(1+\imath )}{\delta }$ \footnote{
We have ignored the reflected travelling waves considering the top
plate to be at infinity. This is because the distance between the
plates is much larger compared to the boundary layer.} . The
frictional force acting on the sphere due to the velocity gradient
is in the direction of the liquid motion. The force per unit area is
given by
\begin{eqnarray}
  S &=& \eta \frac{{\partial v}}{{\partial z}}=- \frac{{\sqrt 2 }}{\delta }\eta v_0 e^{ - i\left( {\omega t + \frac{\pi }{4}} \right)} e^{\left( {i - 1} \right)\frac{z}{\delta
    }}.
\end{eqnarray}
We obtain the total frictional force ($F_s$)  acting on the sphere
by computing the surface integral.
\begin{equation}
F_s  = 2\pi \eta av_0 e^{ - \left( {\frac{{h + a}}{\delta }}
\right)} (1 - e^{\frac{{2a}}{\delta }} )e^{ - i\omega t},
\end{equation}
where $h$ is the height of the sphere  from the glass plate.

 The forces experienced by the trapped particle \ at a height $h$
from the glass plate are (i) the \ spring like force $-k_{op}x$,
exerted by the tweezer; (ii) the viscous drag force $-6\pi \eta
a(\dot x-u)$,where $\dot x$ is the velocity of the sphere and $u$
the velocity of the liquid in the vicinity of the
sphere.(ii)hydrodynamic force $F_h$  and (iv) a frictional force
($F_{plate}$) due to the interaction with the plate.

Thus, the equation of motion of the sphere is
\begin{equation}
m_{eff} \ddot{x}+6\pi \eta a \dot x + k_{op} x = F_{h}  +
F_{plate}+6\pi \eta a u,
\end {equation}
where $m_{eff}$ is the effective mass of the particle. We introduce
$m_{eff}$ in place of actual mass of the particle to account for the
effects of frequency, hydrodynamic interactions, and presence of
wall to the Stoke's drag.
The force acting on the sphere due to the
glass plate is mainly electrostatic in origin and hence strongly
dependent on the sphere-plate separation. When the sphere is far
away from the plate and frequency of oscillations are low, $F_h$ is
the frictional shear force $F_s$. However, in cases where the sphere
is within the stokes boundary layer and the length scale of the
vortex shedding is greater than the plate-sphere separation one
expects flow field around the particle to be altered by the presence
of bottom plate and the inertial effects to differ from those for a
particle in an unbounded flow \cite{FISCHER_oscillatory,Sumer}.Thus,
for the sphere near the bottom plate the hydrodynamic force will
differ from the simple minded frictional shear force.

\section{Experimental Results}

We will first discuss the results which show transition of the
motion of the sphere dominated by inertial contact with the plate to
the regime where the motion of the sphere is mainly driven by the
liquid velocity. Top panel of Fig. \ref{fig:fig2} shows the motion
of the sphere (solid line) suspended in water and the corresponding
drive signal to the $y$-direction of the piezo (dotted line).  The
data plotted is for a constant driving frequency (1Hz) and constant
drive amplitude(1$\mu m$ ) for various sphere-plate separation ($h$)
\footnote{The heights have been calculated from the $z$ displacement
of the piezo.}. The drive signal shown in the top panel is only a
guide to the eye and the value on the y axis is not a reflection of
its absolute magnitude. When $h$ ($\sim$ 0.1 $\mu$ m) is small, the
response of the sphere is in phase with the drive signal. As the
plate-sphere separation is increased the motion of the sphere
develops a phase lag with respect to the plate. This is shown in the
bottom panel of Fig. \ref{fig:fig2} where the histogram of the phase
of the motion of the sphere(averaged over 3 seconds) is plotted as a
function of the sphere-plate separation.`` Phase '' of the motion of
the sphere is defined by the phase of the locked in signal with
respect to the driving signal. The phase of the sphere's motion
confirms the two regimes-inertia coupled (in-phase) and velocity
coupled (finite phase lag). In between the two extremes of in phase
and $\sim$ $\pi /2$ out of phase motion, the histogram of the phase
of the motion of the sphere shows a broad distribution \footnote {On
carefully inspecting the displacement of the sphere one finds the
sphere to show slip-stick behavior.}.

We now present our results which measures the effect of large drive
frequencies on the motion of the sphere for appreciably large values
of $h$. In this case the sphere executes a $\pi /2$ out of phase
motion with respect to the motion of the bottom plate for low drive
frequencies. This confirms that the motion of the sphere is
predominantly velocity coupled. We have used two aqueous mediums of
suspension, namely water($\eta = 1mPas$) and glycerol($\eta=
760mPas$) for this study. Left panel of Fig. \ref{fig:fig3} shows
response of the sphere(solid spheres joined by line) suspended in
glycerol as the frequency of the oscillation of the plate is varied
when the sphere-plate separation is constant at about $\sim$ 1.3
$\mu$m (sphere is far away from the plate).The drive amplitude of
the plate was kept constant at $0.1 \mu m$. The trace of the
velocity of the liquid is shown by a solid line. Note that the phase
of the velocity of the liquid is $\pi /2$ phase shifted with respect
to that of the bottom plate.\footnote{$y_{p}= y_{p0}sin(\omega t)$,
$\dot y_{p}=y_{p0}\omega sin(\omega t+\pi /2)$, where $\dot y_p$ is
the velocity of the plate and assuming no-slip boundary condition is
also the velocity of the liquid.} It can be seen that at 2Hz the
sphere moves almost in phase with the velocity of the liquid.
However, as the drive frequency is increased the sphere develops a
$\sim \pi /2$ phase difference for drive frequency of 11Hz and $\sim
\pi$ phase difference for a drive frequency of 40 Hz. The
corresponding average phase is shown by solid circles joined by line
in top panel of Fig.\ref{fig:fig4}. The inset of the top panel in
fig. \ref{fig:fig4} shows the amplitude of the locked in signal.

The right panel of  Fig. \ref{fig:fig3} shows response of the
sphere(solid spheres joined by line) suspended in water as the
frequency of the oscillation of the plate is varied when the
sphere-plate separation is constant at about $\sim$ 1.3 $\mu$m
(sphere is far away from the plate).The drive amplitude of the plate
was kept constant at $1 \mu m$. The sphere's motion at the driving
frequency of 2Hz is purely sinusoidal and monochromatic, with a
dominant frequency of 2Hz. At this frequency the motion of the
sphere is almost in phase with the velocity of the liquid. However
as the frequency is increased, the motion of the sphere develops a
phase lag with respect to the velocity of the liquid. This is also
accompained by a large distortion of the waveform of the motion of
the sphere. It is interesting to note that such distortions are
absent for the case of glycerol.The corresponding average phase is
shown by solid circles joined by line in bottom panel of
Fig.\ref{fig:fig4}. The inset of the bottom panel in fig.
\ref{fig:fig4} shows the amplitude of the locked in signal.

In the analysis of the phase as shown in fig.\ref{fig:fig4} we have
shifted the phase of the drive signal by $\pi/2$ to account for the
velocity contribution. That is, the phase is plotted with respect to
the velocity of the liquid. As the frequency is increased from 2Hz
to 60Hz the phase decreases monotonically from zero  to $-\pi$.
Since the sphere is far away from the plate we ignore $F_{plate}$.
We can then consider this system as a forced damped harmonic
oscillator. The equation of motion of a forced damped harmonic
oscillator is $ \ddot x + \gamma \dot x + \omega _0^2 x = \frac{{F_0
}}{m}\cos \omega t $ which has a steady state solution $x= A \cos
\omega t$ where $A = \frac{{F_0 }}{m}\frac{1}{{\left[ {(\omega _0^2
- \omega ^2 )^2 + (\omega \gamma )^2 } \right]^{1/2} }} $ and $ \phi
= \arctan \left( {\frac{{\gamma \omega }}{{\omega ^2 - \omega _0^2
}}} \right) $. Comparing it with eq no 3 we see that in our case, $
\gamma = \frac {6\pi\eta a} {m_{eff}},
\omega_0^2=\frac{k_{op}}{m_{eff}}$.

Figure \ref{fig:fig4} also shows the fit (open squares joined by
line) to Eqn., $\phi = \arctan \left( {\frac{{\gamma \omega
}}{{\omega ^2 - \omega _0^2 }}} \right) $ to the experimental data.
From the fit we obtain $\gamma=150$ and $\omega_0=11Hz$ for glycerol
and $ \gamma=220$ and $\omega_0=20Hz$ for water. This implies
implies $m_{eff}=0.095$x$10^{-6}$Kg for the case of glycerol and
$m_{eff}=0.085$x$10^{-9}$Kg for water. The value of $k_{op}=m_{eff}
\omega_0^2$ obtained from the fit parameters for glycerol is
5.96x$10^{-4}$N/m which agrees well with that obtained from the
equilibrium positional fluctuations of the
sphere(2.7x$10^{-4}$N/m).The value of $k_{op}=m_{eff} \omega_0^2$
obtained from the fit parameters for water is 13.4x$10^{-7}$N/m
which agrees well with that obtained from the equilibrium positional
fluctuations of the sphere(11.4x$10^{-7}$N/m). The quality of the
fit improves if  $m_{eff}$ is considered as function of effective
mass. But in that case the equation of motion should be solved using
perturbation theory.

Notice that $m_{eff}$ is much larger than the true mass of the
bead($\sim$ 4.18x$10^{-15}$Kg). We can understand this by saying
that it is not just the sphere but also the fluid around it which is
behaving like an oscillator.

\section{Discussion}

The relevant length scales perpendicular to direction of the flow,
that is, in the z-direction are the particle diameter,($2 \mu m$),
and the stokes boundary layer, $\delta =\sqrt{\frac{2\eta
}{\omega\rho }}$. In frequency range,$1Hz \ldots 60 Hz$, covered in
our experiments $\delta=13mm\ldots 1.6mm$ for glycerol and
$\delta=0.5mm\ldots 0.07mm$ for water. In the direction parallel to
the flow, one of the relevant length scales is again the particle
diameter ($2 \mu m$) and the other is associated with vortex
formation, shedding and potential interactions. For a sinusoidal
motion, the length scale associated with vortex formation is roughly
equal to amplitude of the plate motion \cite{FISCHER_oscillatory}
which in our experiments is about 0.1$\mu m$ for glycerol and 1$\mu
m$ for water.

The motion of the sphere comprises of two kinds of degrees of
freedom- translation and rotational. Translation motion arises
because whole fluid around it translates when the bottom plate is
sinusoidally driven. Since the sphere is trapped and is subjected to
shear stresses there will be a finite torque leading to the rotation
of the sphere about the $x$ axis with an angular velocity$ {\bf
\Omega } = \left( {\Omega,0,0} \right) $. No-slip boundary condition
on the sphere's surface ensures that the curl of the velocity of the
liquid is non zero around the surface of the sphere. The Navier
Stokes equation which defines the flow of the liquid about the
sphere is given by $ \nabla ^2 {\bf q} - \nabla p = \frac{\rho
}{\eta }\left( {{\bf q}.\nabla } \right){\bf q} + \frac{\rho }{\eta
}\frac{{\partial {\bf q}}}{{\partial t}}, $ where $\bf
q=(q_1,q_2,q_3)$ is the velocity of the fluid. The origin of the
cartesian coordinate is taken to coincide with the center of the
sphere. The boundary conditions are ${\bf q} \to 0$ when $z$ tends
to infinity, $ \bf q = \bf \Omega \times \bf r$ for the flow on the
sphere and ${\bf q} \to $ velocity of the plate for the flow on it,
here $\bf r = (x,y,z)$.This will lead to vortex formation which
decays with a length scale of the boundary layer thickness. It is
noteworthy that the stokes boundary layer is of the order of at
least tens of microns for both glycerol and water in the frequency
range covered in our experiments. Since the sphere is within the
boundary layer, one expects considerable potential interactions
between the rotational flow of the liquid and the oscillating plate.
This scenario could possibly result in  an inertial coupling between
the plate and the sphere. This in turn could increase the effective
mass of the sphere and hence exhibit the observed phase behavior.

\section{Conclusion}
To our knowledge,there has been no experimental/computational fluid
dynamics simulations of a  particle, rotating as well as
translating, in an oscillatory flow. Fischer \textit {et al} have
performed calculation of lift and drag forces on a stationary sphere
subjected to a pressure driven oscillatory flow. Our results are the
first direct measurements of the drag forces acting on a rotating
sphere subjected to an oscillatory motion. We find that the phase of
the motion of the sphere with respect to the drive is a sensitive
tool to study its dynamics. We have been able to explain our data in
terms of a damped harmonic oscillator. The effective mass that comes
out of the calculations is orders of magnitude greater than the bare
mass of the sphere. This highlights the importance of the role of
inertia in an otherwise viscosity dominated flow. We believe that
optical tweezers will be a effective tool to address fluid dynamics
problems at small length scales.

\clearpage

 \newpage
\begin{figure}
    \centering
    \centerline{\psfig{figure=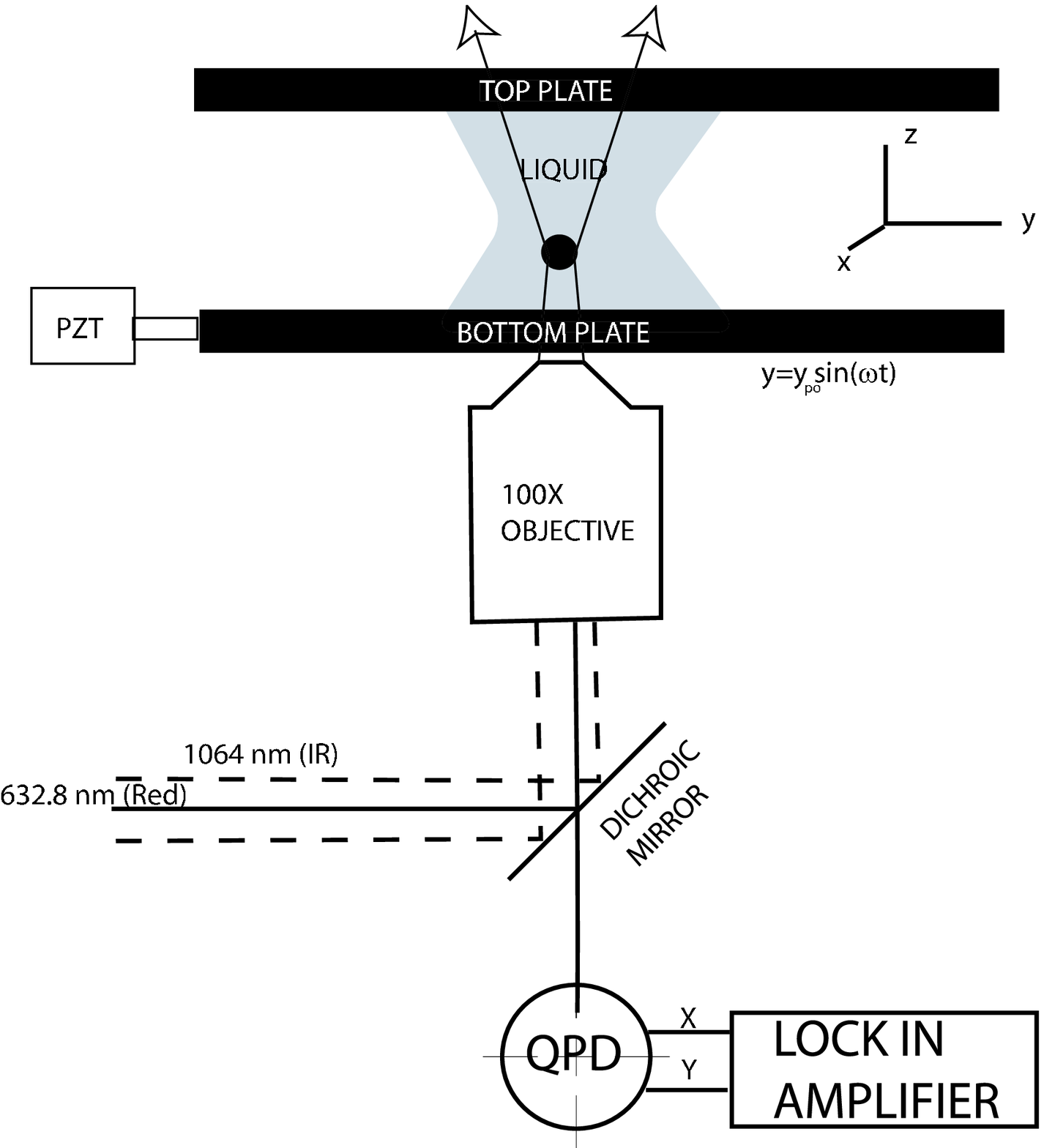,width=12cm} }

    \caption{Schematic of the experimental setup. Liquid containing 2$\mu m$ colloidal particles is held between two glass plates separated by 2mm. The particle is optically trapped at a height $h$ from the bottom plate. The lower plate is subjected to an oscillatory  motion   $y_{p}=y_{po}\sin (\omega t)$, where $y_{p}$ is
   the instantaneous displacement of the glass plate, $\omega$ is angular velocity and $y_{po}$ is the amplitude. The top plate is kept fixed. }
    \label{fig:fig1}
\end{figure}

 \newpage
\begin{figure}

    \centering
    \centerline{\psfig{figure=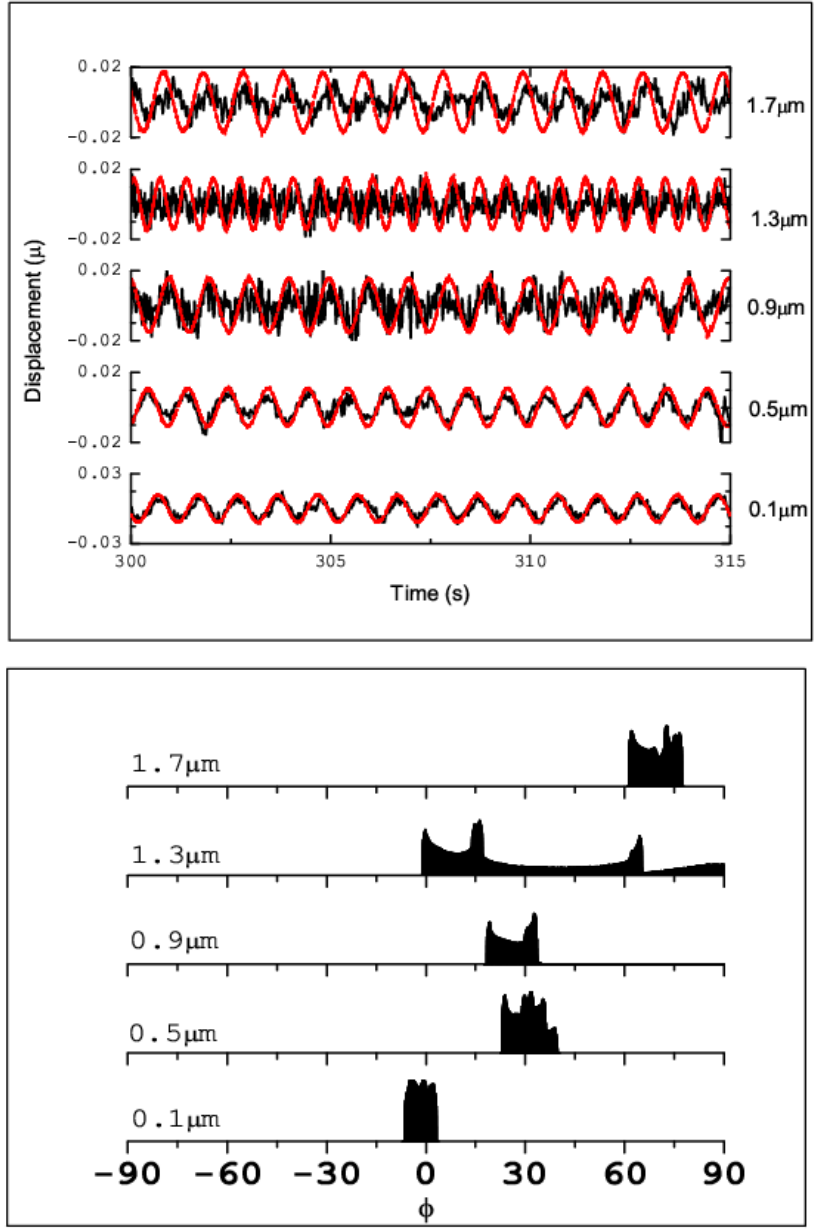,width=12 cm} }
    \caption{ Top : The motion of the sphere(solid line) as a function of the sphere-plate separation: (a)$\sim$ 0.1 $\mu$m (b)$\sim$ 0.5 $\mu$m (c)$\sim$ 0.9 $\mu$m
(d) $\sim$ 1.3 $\mu$m. (e)$\sim$1.7 $\mu$m at a fixed driving
frequency of 1Hz and amplitude of 1$\mu m$. The trace showing the
motion  of the plate (dotted line) is a guide to the eye, and the
corresponding $y$ axis is not a reflection of its absolute
amplitude.
            Bottom : The corresponding histograms of the phase of the locked in signal as a
function of the sphere-plate separation. The y axis for the
histograms is in logarithmic scale. }
    \label{fig:fig2}
\end{figure}

 \newpage
\begin{figure}
    \centering
    \centerline{\epsfig{figure=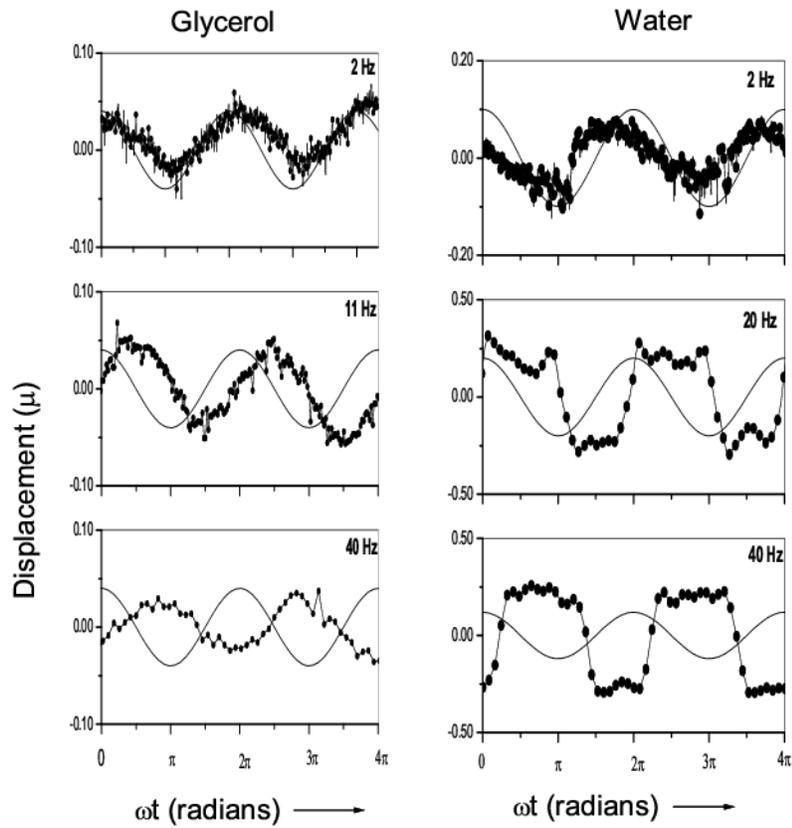,width=12 cm} }
    \caption{   Displacement of the sphere (solid circles joined by lines) in response to the velocity of the liquid (Solid line) for few select frequencies. The frequencies are shown by the side of each trace.  Left: The data shown is for glycerol, Right : The data shown is for water. The trace showing the velocity of the liquid is a guide to the eye, and the corresponding $y$ axis is not a reflection of its absolute amplitude.   }
    \label{fig:fig3}
\end{figure}

 \newpage

\begin{figure}
    \centering
    \centerline{\psfig{figure=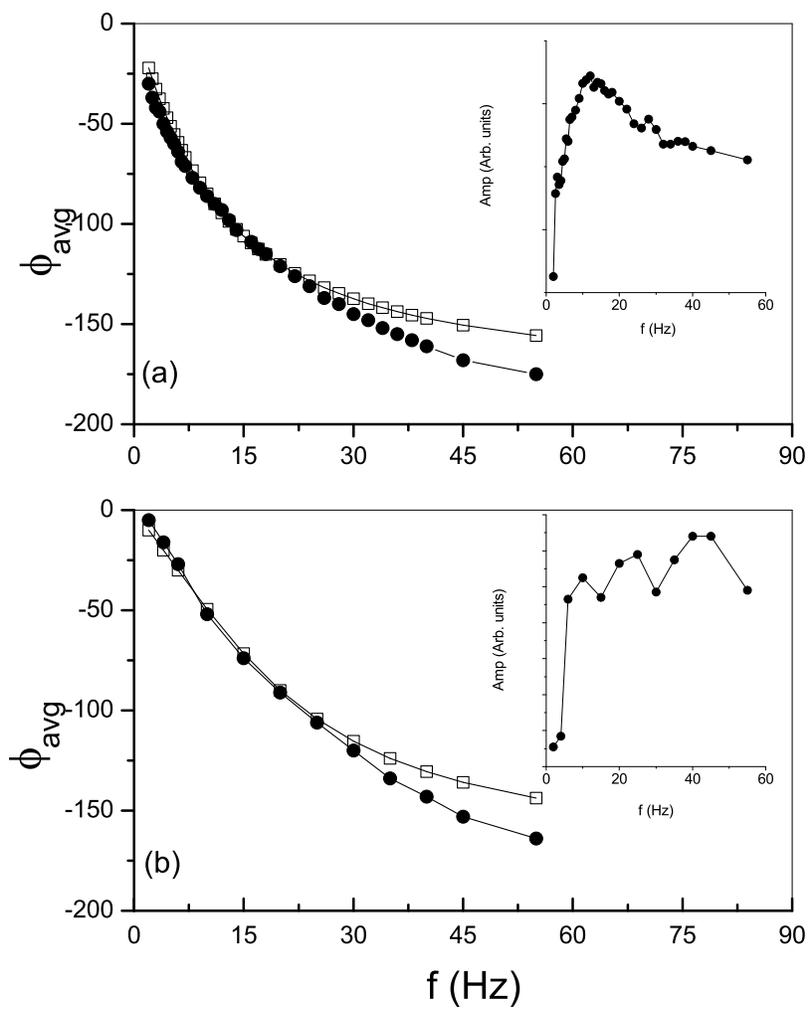,width=12cm} }
    \caption{The variation of average phase with respect to the velocity of the liquid. The inset shows the amplitude of the locked in signal. Top: The data shown is for glycerol.
Bottom: the data shown is for water   }
    \label{fig:fig4}
\end{figure}

 \newpage

\end{document}